# High-Temperature Annealing of TiO₂ Nanotube Membranes for Efficient Dye-Sensitized Solar Cells

by Fatemeh Mohammadpour,[1] Marco Altomare,[2] Seulgi So,[2] Kiyoung Lee,[2] Mohamed Mokhtar,[3] Abdelmohsen Alshehri,[3] Shaeel A. Al-Thabaiti,[3] and Patrik Schmuki[2,3]*

[1] Department of Physics, Faculty of Science, University of Shiraz, Eram Street, 71454, Shiraz (Iran)

[2] Department of Materials Science and Engineering, WW4-LKO, University of Erlangen-Nuremberg Martensstrasse 7, 91058 Erlangen (Germany)

[3] Department of Chemistry, King Abdulaziz University Jeddah (Saudi Arabia)

* Corresponding author. E-mail: schmuki@ww.uni-erlangen.de, Tel.: +49-9131-852-7575, Fax: +49-9131-852-7582

**A B S T R A C T**

We fabricate photo-anodes by transferring anodic $TiO_2$ nanotube membranes in tube-top-down configuration on FTO glass, and use them for constructing frontside illuminated dye-sensitized solar cells. Prior to solar cell construction, the tube-based photo-anodes are crystallized at different temperatures (400-800°C), and the effects of tube electron transport properties on the photovoltaic performance of the solar cells are investigated. We show that improved solar cell efficiencies (up to *ca*. 8.0%) can be reached by high-temperature treatment of the tube membranes. Consistently with electron transport time measurements, remarkably enhanced electron mobility is enabled when tube membranes are crystallized at 600°C.

**Introduction**

In the last decades, $TiO_2$-based dye-sensitized solar cells (DSSCs) have attracted a great deal of attention, being cost-effective and delivering higher efficiency compared to solar cells fabricated from other semiconductor metal oxides. Grätzel-type solar cells [1,2] were shown to reach power conversion efficiencies even higher that 12% [3], this by using dye-sensitized photo-anodes constructed from nanoparticulate $TiO_2$ films.

In view of enhancing the cell efficiency, aside from improving the light absorption ability by the dye-sensitized photo-anode and the electron injection from the excited dye into the semiconductor scaffold, other essential aspects are the electron transport through the photo-anode and the electron collection at the oxide scaffold/electrode interface.

The electron transport through nanoparticulate photo-anodes is intrinsically limited by the "random walk" through the $TiO_2$ network, and the efficiency of the solar cell can thereby be detrimentally affected [4–6]. Thus, photo-anodes based on one-dimensional (1D) nanostructures, like nanowire [7,8], nanorod [9,10] and nanotube [11–16] arrays, being expected to provide a directional (*i.e.*, faster) charge transport [14,15], have been investigated as promising candidates in view of building more efficient DSSCs.

Particularly, anodic $TiO_2$ nanotubes (NTs) were intensively explored for solar cell applications, not only because of their highly ordered 1D structure, but also for their low-cost and straightforward fabrication [17,18], which is simply based on the anodization of a piece of Ti metal under self-organizing electrochemical conditions [19].

Remarkable leaps in nanotube-based solar cell efficiencies could be accomplished by adopting a front-side illumination configuration for the device (Scheme 1(a)), this to avoid the losses of light by absorption in the electrolyte and in the Pt-coated counter electrode. Different techniques have been developed, that allow for complete detachment of entire anodic tube arrays (*i.e.*, free-standing membranes) from the titanium substrate, that can then be transferred onto optically transparent conductive glass slides (Scheme 1) to fabricate photo-anodes for frontside irradiated solar cells [20–23].

Another key for improving the charge transport and collection kinetic lies in a suitable crystallization process [13,14,24–28]. Some recent works illustrate different low-temperature crystallization process (*i.e.*, water annealing [29,30], water vapor [31] and hydrothermal [32] treatments and photo-induced crystallization [33,34]) as possible alternatives to the conventional heat treatment. However, these methods may lead either to only partial crystallization of the $TiO_2$ nanostructures (*i.e.*, high density of recombination states) or to substantial loss of their tubular morphology [30,32]. Conversely, a complete crystallization into anatase phase, which is essential to enable efficient charge carrier separation and electronic transport in photo-electrochemical applications, can be easily obtained by classic thermal annealing [30].

It was also recently shown that an optimized annealing profile leads to considerable improvement of the tube electron transport properties [35]. Hence, on the basis of these results, one would expect not only the annealing profile but also the tube crystallization temperature to markedly affect the solar cell efficiency [27,28].

In this work we fabricate front-side illuminated DSSCs from anodic $TiO_2$ nanotube membranes. The tube arrays used for building the photo-anodes are lifted-off from the Ti substrates by a simple re-anodization method [21], and are annealed at different temperatures in the 400-800°C range prior to solar cell construction. We show that faster electron transport through the $TiO_2$ scaffolds is enabled when tube membranes undergo crystallization at relatively high-temperature, *i.e.*, 600-700°C, this leading to significant improvement of the solar cell efficiency up to *ca.* 8%.

**Experimental**

Highly ordered TiO$_2$ nanotube arrays were prepared by potentiostatic electrochemical anodization of Ti foils (0.125 mm thick, 99.7 % purity, Advent) in a two-electrode electrochemical cell, by using a Pt foil as counter electrode (the applied bias was provided by a Voltcraft VLP2403pro power source). Before anodization, the Ti foils were cleaned in acetone, ethanol and deionized water by sonication, and then dried in a N$_2$ stream.

The anodization experiments were performed at 60 V, in an ethylene glycol-based electrolyte with 3 vol% of deionized water and 0.15 M of NH$_4$F for 1 h, this to reach a tube length of 20 μm. For the lift-off of the tube membranes we followed a simple re-anodization approach [21,22]: the as-prepared samples were annealed at 350°C in air for 1 h, this leading to their crystallization, and then were anodized again in identical conditions, so to form a second tube layer beneath the crystalline tube array. The underlying amorphous film can be preferentially dissolved by immersion in 0.07 M HF aqueous solution thermostated at 30°C, while the crystalline tube layer, being more stable in such environment, thereby detaches as entire self-standing nanotube membrane.

The tube membranes were left in air for drying, sandwiched between two quartz slides (this step avoids curling and cracking of the tube films during crystallization) and annealed in air, for 1 h, at different temperatures in the 400-800°C range, with a heating/cooling rate of 30°C min$^{-1}$ using a Rapid Thermal Annealer (Jipelec JetFirst100).

After crystallization, the membranes were transferred in tube-top-down configuration [23] (Scheme 1) on FTO slides (7 Ω m$^{-2}$) that were previously coated with *ca.* 2 μm-thick TiO$_2$ nanoparticle film (Ti-Nanoxide HT, Solaronix) deposited by doctor-blade method.

After drying in air for *ca.* 30 min, the prepared photo-anodes underwent a further annealing step, in air for 1h, in order to grant good adhesion, mechanical stability and established good electric contact between the tubes, the TiO$_2$ nanoparticle layer and the conductive substrate (*i.e.*, FTO). For this, preliminary screening showed a temperature of 400°C to be adequate for achieving good mechanical robustness and cell functionality.

The as-prepared photo-anodes were then immersed in a 300 μM dye solution (D719, Everlight, Taiwan) at 40°C for 24 h. After dye-sensitization, the samples were rinsed with acetonitrile to wash off the non-chemisorbed dye and then were dried in a N$_2$ stream.

The dye-sensitized photo-anodes were sandwiched with a Pt coated FTO glass as counter electrode, by using a hot-melt spacer (25 μm, Surlyn, Dupont). The fabrication of the DSSCs was completed by introducing the electrolyte (Io-li-tec, ES-0004) in the interspace between the two electrodes.

For the morphological characterization of the TiO$_2$ nanotube membrane-based photo-anodes, a field-emission scanning electron microscope (FE-SEM Hitachi S4800) was used. X-ray diffraction (XRD) analysis was performed with an X'pert Philips MPD diffractometer, equipped with a Panalytical X´celerator detector and using graphite monochromized Cu K$\alpha$ radiation ($\lambda$ = 1.54056 Å).

The current-voltage characteristics (*i.e.*, J-V curves) of the solar cells were measured under AM 1.5 simulated solar light illumination provided by a solar simulator (300 W Xe light with optical filter, Solarlight), and by applying an external bias to the cell (from -50 mV up to +900 mV) and measuring the generated photocurrent with a Keithley 2420 digital source meter. Step size and holding time were 23.75 mV and 100 ms, respectively. The active area was defined by the 0.2 cm$^2$-sized opening of the Surlyn seal and a scattering background was used. Typically, 3 to 5 cells were produced for each crystallization temperature, and the measured photovoltaic parameters resulted reproducible, with an average scattering within the 5%.

Intensity modulated photovoltage and photocurrent spectroscopy (IMVS and IMPS) measurements were carried out using modulated light (10% modulation depth) from a high power green LED ($\lambda$ = 530 nm). The modulation frequency was controlled by a frequency response analyzer (FRA, Zahner). The light intensity incident on the cell was measured with a calibrated Si photodiode. In general, IMPS and IMVS data were observed to be consistent, and with an average scattering within the 10%.

The dye-loading of the photo-anodes was determined by immersing the dye-sensitized photo-anodes in 5 ml of a 10 mM NaOH aqueous solution for 30 minutes. Then, the absorption of the solutions was measured by UV-Vis spectrophotometer (Lambda XLS+, Perkin Elmer).

**Results and discussion**

Examples of the anodic $TiO_2$ nanotube arrays fabricated in this work are shown in Scheme 1(b) and Fig. 1. These tubes grow highly aligned and show mean diameter of ~ 100 nm and length of 20 μm.

As outlined in the experimental section, the membranes were sandwiched between two quartz slides after lift-off and crystallized at different temperatures in the 400-800°C range. The SEM images in Fig. 1 show that the morphology of the tubes is clearly affected by the thermal treatment, particularly when crystallization is carried out at T ≥ 500-600°C.

While the crystallization at 400°C does not induce any evident modification in the tubes appearance compared to the as-formed layers, a thermal treatment at 500°C leads to clearly visible grain boundaries and "pinholes" along the tube walls [36]. More visible changes occur at 600-700°C: the walls of the tubes undergo thickening, more pinholes form and the mean size of $TiO_2$ crystallites increases (see XRD data below).

At 800°C we observed severe coarsening of the tube walls, which completely merged with the adjacent ones, thereby forming an interconnected three-dimensional matrix [37]. Although the conversion of the tubular structures into a porous matrix gradually takes place at 500-600°C, it is rapidly achieved at T > 700°C (this is clear from the top view SEM micrograph in Fig. 1).

In addition to the structural changes at the nanoscale, it is commonly observed that the thermal annealing generates microscopic cracks in the $TiO_2$ tube layers [38–40]. However, we found that the tube layers could withstand the crystallization step by retaining a good mechanical stability, so that these membranes could be processed for assembling DSSCs even when annealed at 800°C (*i.e.*, complete membrane cracking was not observed even at such high temperature).

Fig. 2(a) shows J-V curves of different tube membrane-based solar cells, while table in Fig. 2(b) summarizes the photovoltaic characteristics (measurements were performed under AM 1.5 simulated solar light illumination). As evident from these data, the annealing temperature strongly influences the performance of the tube photo-anodes. A most optimized condition that led to our highest solar cell efficiency of 8.05% is a crystallization of the NT membranes at 600°C.

In the light of such results, XRD patterns of the photo-anodes annealed at different T were collected (Fig. 2(c)). As expected, and well in line with previous reports [37], the thermal treatment led to conversion of amorphous (as-formed) anodic oxide into $TiO_2$ anatase NT membranes, this regardless of the annealing temperature. In other words, no rutile formation could be detected, this more likely owing to the absence of the Ti metal substrate that typically triggers the anatase-into-rutile conversion in the tube walls by rutile seeding ascribed to Ti thermal oxidation [27,36,41] (please note that the rutile formation would be detrimental to the solar cell efficiency since this $TiO_2$ polymorph typically shows relatively poor electronic properties [4,15]).

Although no remarkable difference in phase composition could be seen from the XRD patterns of differently annealed membranes, a gradual increase of intensity of main anatase peak at 25.3° was observed (this is associated to higher crystallinity degree for tubes annealed at higher T). Also, the XRD data (processed by applying the Scherrer equation to the (101) anatase reflection) reveal that the mean crystallite size varied from *ca.* 23.8 to 26.5 nm, when the annealing temperature was increased from 400 up to 700-800°C (these data are well in line with previous works [37,42,32]).

In order to investigate charge recombination and transport properties of the tube membranes used for fabricating solar cells, we performed IMVS and IMPS measurements. IMVS data (compiled in Fig. 2(d)) show that recombination time decreased by increasing the annealing temperature (this trend was observed independently of the photon flux). Hence, the improved photovoltaic performance of tube membranes annealed at relatively high T seems not to find explanation in charge recombination data.

IMPS results (Fig. 3) showed instead that the electron transport in the tubes becomes remarkably faster particularly when membranes are crystallized at 600°C, being such thermal treatment also optimal, as anticipated, for solar cell performance (Fig. 2(a)-(b)). One can therefore assume that when the annealing temperature is increased up to 600-700°C, the charge transport properties of the tube photo-anodes are improved, and this may prevail over the slight drop of recombination time observed at similar temperatures (Fig. 2(d)).

More precisely, solar cells fabricated from photo-anodes annealed at 600°C, besides delivering the highest power conversion efficiency when compared to other devices, also showed the highest $J_{SC}$ and $V_{OC}$ (17.56 mA cm$^{-2}$ and 0.80 V, respectively). Typically, a key factor that largely affects $J_{SC}$ is the dye-loading, that is, the larger the amount of dye on the $TiO_2$ scaffold, the more efficient the light absorption and consequently the higher the resulting current density [22]. An effective approach for improving the efficiency of nanotube-based solar cell implies in fact the decoration of tubes with $TiO_2$ nanoparticles which, by increasing the surface area of the semiconductor material, improve thereby the dye-loading ability of the photo-anode [43].

However, it should be noted that in the present study a clear drop of dye-loading was observed when increasing the annealing temperature, which is a direct consequence of crystallite growth, sintering of the tubular structures and loss of surface area.

BET analysis was carried out for differently annealed nanotubes in order to determine their specific surface area in relation to the crystallization temperature. Values in the range of *ca*. 30-50 $m^2$ $g^{-1}$ were determined, which are in line with results reported by previous works [29,44]. However, the BET data showed no specific trend in relation to the different annealing temperatures. In general, we found that a 20 μm-thick tube membrane has an average weight of ~ 1.5 mg, while for reliable BET analysis amounts of material of *ca*. 100 mg are needed (when the expected BET area is of some tens of $m^2$ $g^{-1}$). Therefore, a certain lack of accuracy may arise from such discrepancy.

Fig. 4(a) shows that the drop of dye-loading occurs to the largest extent when annealing the membranes in the 500-700°C range. Even more, the tube scaffolds crystallized at 600°C showed a drop of the dye-loading of around 25% compared to the tubes annealed at 400°C, this although the former gave best photovoltaic performances. Thus, within the parameter frame, dye-loading and specific surface area of the tube scaffolds seem not be key factors.

Interestingly, we noticed that when plotting the photo-voltaic data *vs.* the crystallization temperature as illustrated in Fig. 4(b), both the trend of $V_{OC}$ and that of $J_{SC}$ data exhibited a maximum for the cells fabricated from photo-anodes crystallized at 600°C. Precisely, while the trend of $J_{SC}$ slightly scatters at T ≥ 700°C (at 700°C not only we measured the smallest $J_{SC}$ but also the shortest recombination time), that of $V_{OC}$ matched rather well the trend of cell efficiency.

Even more, such trend was mirrored by IMPS results (Fig. 3), so that one may interpret these results assuming that annealing at 600-700°C leads to higher degree of crystallinity for the $TiO_2$ scaffolds, which in turn reduces trap state density in the tubes and therefore improves significantly the charge transport properties of the photo-anodes. Also remarkable is that the electron transport was enhanced, and charge collection kinetic overall improved by annealing the tube membranes at 600°C, this although the thermal treatment at such temperature simultaneously led to the drop of both recombination time and dye-loading.

Perspective wise, even higher efficiencies for solar cells fabricated from tube membranes could be achieved if not only charge transport, but also recombination time and dye-loading would be improved as well, and recent studies showed the feasibility of these strategies [35,43].

**Conclusion**

In this work we fabricated frontside illuminated dye-sensitized solar cells from anodic $TiO_2$ nanotube membranes, and particularly investigated the effect of crystallization temperature of the tube-based photo-anodes on the solar cell performance. We showed that when membranes are annealed at 600-700°C, improved electron transport properties of the $TiO_2$ scaffolds are enabled which consequently leads to remarkable improvement in solar cell efficiencies, this although such thermal treatment induced crystallite growth, extended tubular structure sintering and worsening of dye-loading ability.

**Acknowledgements**

This project was funded by the Deanship of Scientific Research (DSR), King Abdulaziz University, under grant no. 16-130-36-HiCi. The authors, therefore, acknowledge with thanks DSR technical and financial support. The authors would like to acknowledge the European

Research Council (ERC), the German Research Foundation (Deutsche Forschungsgemeinschaft DFG), and the Erlangen DFG cluster of excellence for financial support. Financial support from the Iran Ministry of Science, Research and Technology is also gratefully acknowledged.

**Figure captions**

**Scheme 1**

(a) Sketch of anodic TiO$_2$ nanotube-based frontside illuminated DSSC. TiO$_2$ tube arrays are detached from Ti substrates by re-anodization approach and transferred onto FTO slides in "tube-top-down" configuration. (b) Cross-sectional SEM image of a photo-anode used for constructing solar cells. As illustrated in (a), a 20 µm-thick self-standing TiO$_2$ nanotube membrane is transferred onto FTO slide that was previously coated with a bladed 2 µm-thick TiO$_2$ nanoparticle film.

**Figure 1**

Cross-sectional and top-view SEM images of TiO$_2$ NT membranes annealed at different temperatures (400-800°C).

**Figure 2**

(a) J-V curves and (b) photovoltaic parameters for DSSCs fabricated from TiO$_2$ nanotube membranes annealed at different temperatures (400-800°C). (c) XRD patterns of membranes annealed at different temperatures (400-800°C, air, 1 h). The patterns were collected after transferring the annealed membranes on FTO glass slides, the membrane transfer being followed by thermal treatment at 400°C to ensure good adhesion of the tube layers and mechanical stability of the photo-anodes. (d) IMVS results (performed under different incident photon fluxes) showing the recombination time of DSSCs fabricated from membranes annealed at different temperatures (400-800°C). Continuous profiles are a guide to the eye.

**Figure 3**

IMPS results (performed under different incident photon fluxes) showing the transport time of DSSCs fabricated from membranes annealed at different temperatures (400-800°C). Continuous profiles are trend lines.

**Figure 4**

(a) Dye-loading, (b) $V_{OC}$, $J_{SC}$ and power conversion efficiencies data plotted *vs.* annealing temperatures (continuous profiles are a guide to the eye).

**Scheme 1**

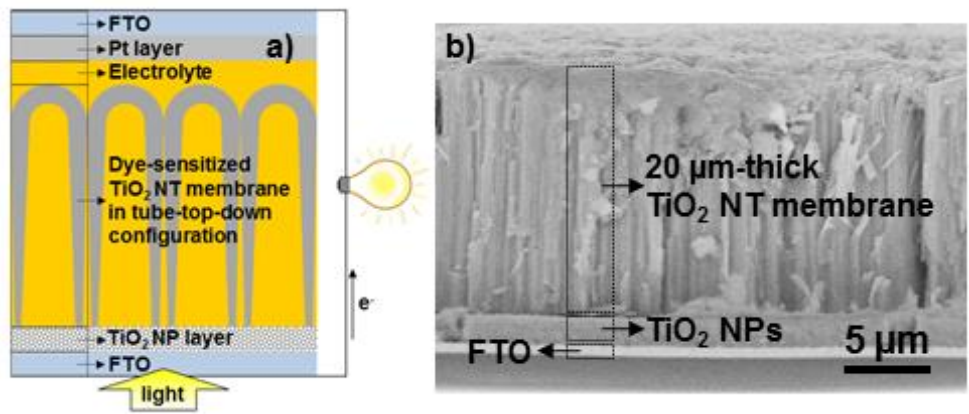

**Figure 1**

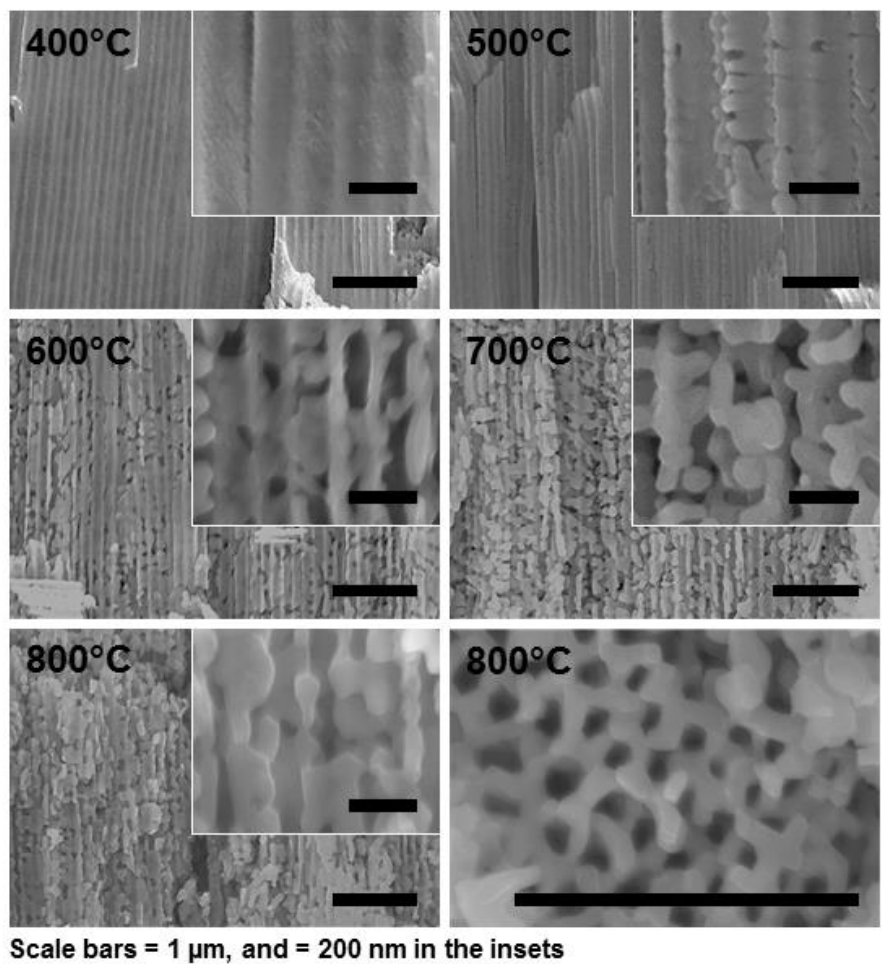

Scale bars = 1 µm, and = 200 nm in the insets

**Figure 2**

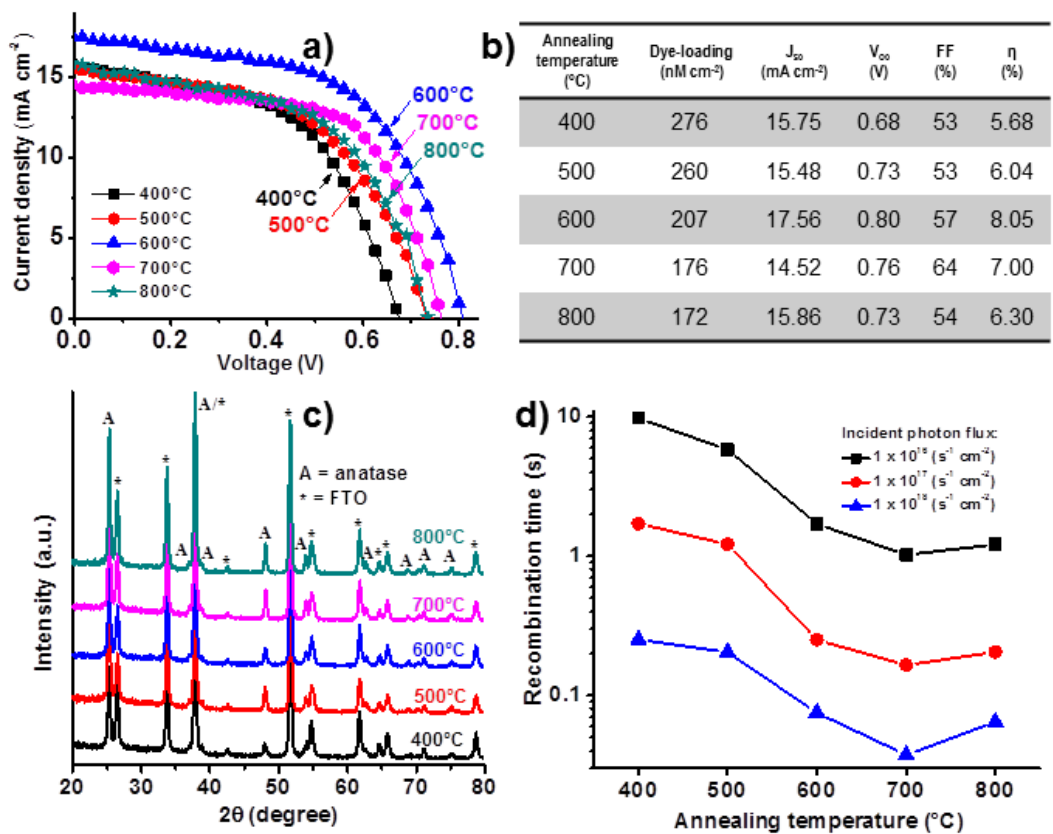

**Figure 3**

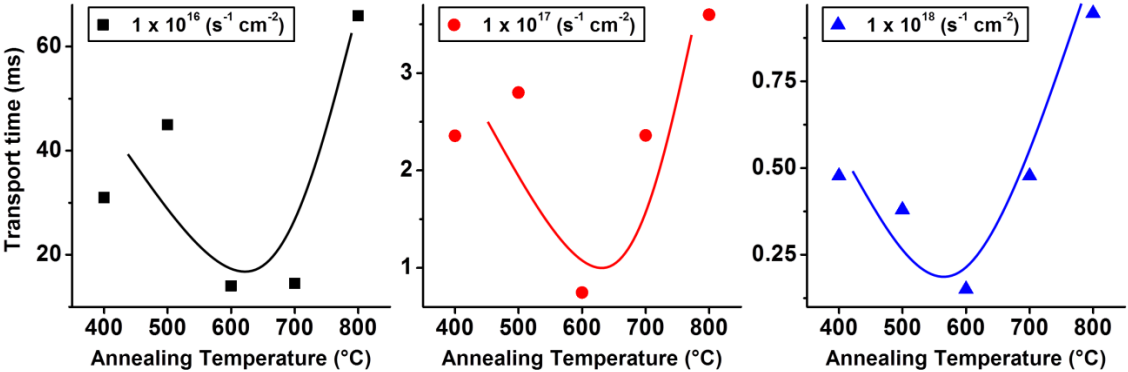

**Figure 4**

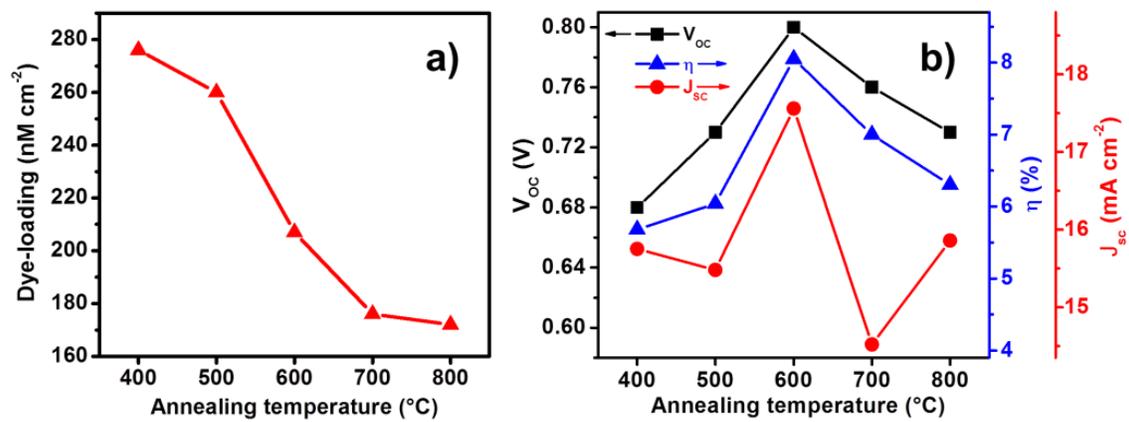